\documentclass[12pt]{article}
\usepackage{epsfig}

\def\beq{\begin{equation}}
\def\eeq#1{\label{#1}\end{equation}}
\def\eeqn{\end{equation}}

\def\beqa{\begin{eqnarray}}
\def\eeqa#1{\label{#1}\end{eqnarray}}
\def\eeqan{\end{eqnarray}}

\let\bar=\overbar

\def\Dslash{\not{\hbox{\kern-4pt $D$}}}
\def\dslash{\not{\hbox{\kern-2pt $\del$}}}

\def\msb{{\bar{\ssstyle M \kern -1pt S}}}

\def\Bbar    {\kern 0.18em\overline{\kern -0.18em B}{}}
\def\Kbar    {\kern 0.18em\overline{\kern -0.18em K}{}}
\def\babar{\mbox{\sl B\hspace{-0.4em} {\small\sl A}\hspace{-0.37em} \sl B\hspace{-0.4em} {\small\sl A\hspace{-0.02em}R}}}

\def\Title#1{\begin{center} { {\bf #1} } \end{center}}

\begin{document}

\Title{\Large Angular Analysis of \mbox{\boldmath$B \to K^* \ell^+ \ell^-$} in 
\mbox{\boldmath$B$}\kern-0.12em 
\large \mbox{\boldmath$A$}\kern-0.12em 
\Large \mbox{\boldmath$B$}\kern-0.12em 
\large \mbox{\boldmath$A\kern-0.12em R$}}

\bigskip\bigskip

%+\addtocontents{toc}{{\it D. Reggiano}}
%+\label{ReggianoStart}

\begin{raggedright}  

{\it Jack L. Ritchie\index{Reggiano, D.}\\
Department of Physics \\
University of Texas at Austin\\
Austin, TX  78712, USA \\
Representing the \babar\ Collaboration}
\bigskip\bigskip
\end{raggedright}

\noindent {Proceedings of CKM 2012, the 7th International Workshop on the CKM Unitarity 
Triangle, University of Cincinnati, USA, 28 September - 2 October 2012}

\section{Introduction}

\begin{figure}[htb]
\begin{center}
\epsfig{file=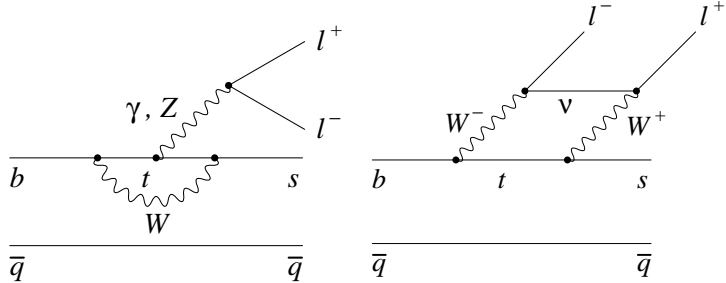,height=1.5in}
\caption{The electroweak penguin (left) and $W$-box (right) diagrams
responsible for $B \to K^{(*)}\ell^+ \ell^-$ decays.}
\label{fig:slldiags}
\end{center}
\end{figure}

Flavor-changing neutral current $B$ meson decays based on the 
process $b \to s \ell^+ \ell^-$ (where $\ell = e$ or $\mu$) , such as 
$B \to K^* \ell^+ \ell^-$, 
provide promising probes for new physics.
The decays proceed through loop diagrams such as those shown in Figure~\ref{fig:slldiags}.

%\begin{figure}[h]
%\centering
%\includegraphics[width=120mm]{figure1.ps}
%\caption{The electroweak penguin (left) and $W$-box (right) diagrams
%responsible for $B \to K^{(*)}\ell^+ \ell^-$ decays.} 
%\label{fig-slldiags}
%\end{figure}

\begin{figure}[htb]
\begin{center}
\epsfig{file=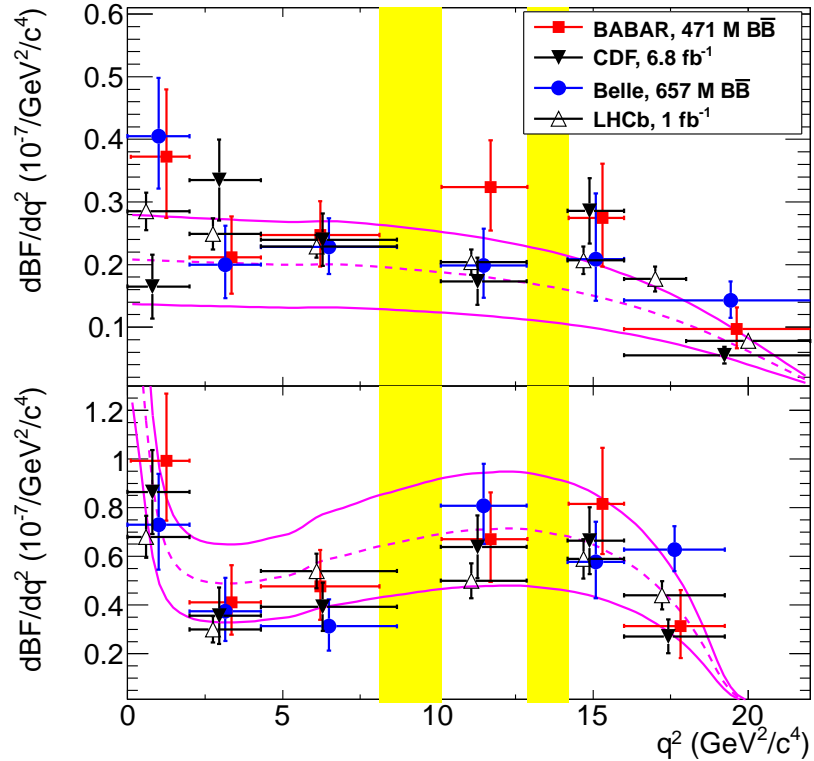,height=4.0in}
\caption{Partial branching fractions versus $q^2$  for
 $B \to K \ell^+ \ell^-$ (top) and $B \to K^* \ell^+ \ell^-$ (bottom)
for \babar\cite{BABARpbfs}, Belle\cite{Belle}, CDF\cite{CDF}, and LHCb\cite{LHCb-Kplus,LHCb-Kstar}.
CDF and LHCb results are for $\mu^+\mu^-$ modes only.  The region between the magenta curves
shows the SM range.\cite{Ali1}  Yellow shading shows
the charmonium exclusion region used by \babar.}
\label{fig:pBFvsq2}
\end{center}
\end{figure}

The theoretical treatment of $b \to s \ell^+ \ell^-$ transitions in the Standard Model (SM) follows
an effective field theory approach in which the Hamiltonian is a sum of terms consisting
CKM factors and Wilson coefficients that multiply 
operators formed from the light quark and lepton fields.  The Wilson coefficients,
obtained by integrating out the heavy particles, characterize the
short-distance physics in these decays.    New physics would
modify the Wilson coefficients by providing new particles inside the loops and may
modify the Hamiltonian by adding additional scalar or pseudoscalar terms.  
To account for QCD effects that mix the operators, so-called effective Wilson
coefficients are defined.
Measurements of $b \to s \ell^+ \ell^-$
decays 
probe the effective Wilson coefficients
$C_7^{\rm eff}$, $C_9^{\rm eff}$, and $C_{10}^{\rm eff}$.\footnote{
$C_7^{\rm eff}$~represents~the~electromagnetic~penguin; 
its~magnitude~is~also~probed~by~$b \to s \gamma$. 
$C_9^{\rm eff}$ and $C_{10}^{\rm eff}$ represent vector and axial-vector components, respectively,
of the $Z$-penguin and $W$-box diagrams.}
However, branching fraction
measurements alone are not sufficient to exploit this opportunity, since they are generally in good
agreement with SM theory predictions, even when measured versus $q^2$ ($= m^2_{\ell \ell}$)
as shown in Figure~\ref{fig:pBFvsq2}.

%A summary of partial branching fraction measurements for all experiments versus $q^2$ for
% $B \to K \ell^+ \ell^-$ and $B \to K^* \ell^+ \ell^-$ is
%shown in Figure~\ref{fig:BFs}.  There is generally good agreement between the measurements and the
%SM theory prediction.

More incisive observables are needed.
Of particular interest is the lepton forward-backward asymmetry, $A_{FB}$, which 
in the SM exhibits a predictable $q^2$ dependence that includes a zero-crossing point
near $q^2 = 4 \, {\rm GeV^2/c^4}$.  Non-SM processes may change the magnitudes and relative
signs of the relevant Wilson coefficients, and induce relative complex phases between them,
and in general may lead to large deviations of $A_{FB}$ versus $q^2$ from its SM expectation.

In this note, \babar's final (but still preliminary) angular analysis of the $B \to K^* \ell^+ \ell^-$
decays is presented, based on a sample of 465 million $B \Bbar$ pairs collected at the $\Upsilon(4S)$
at the SLAC PEP-II B-factory.  This includes measurements of $F_L$, the fraction of longitudinal $K^*$ polarization, and lepton forward-backward asymmetry
%and lepton forward-backward asymmetry\footnote{$A_{FB}$ measures a forward-backward 
%asymmetry in the angle
%of the $\ell^+$ is more often in the same hemisphere  an asymmetry in the distribution
%of the angle between the directions of
%the $\ell^+$ and the $B$ in the $\ell^+ \ell^-$ center-of-mass frame.}
$A_{FB}$ in six $q^2$ bins (as defined in Table~\ref{tab:FLAFB}).
 
\section{Data Analysis}

In this analysis $B \to K \ell^+ \ell^-$ and $B \to K^* \ell^+ \ell^-$ events are reconstructed in nine
distinct submodes: $B^+ \to K^+ e^+ e^-$, $B^+ \to K^+ \mu^+ \mu^-$,
$B^0 \to K^0_S e^+ e^-$, $B^0 \to K^0_S \mu^+ \mu^-$;
$B^+ \to K^+ \pi^0 e^+ e^-$, %$B^+ \to K^+ \pi^0 \mu^+ \mu^-$,
$B^+ \to K^0_S \pi^+ e^+ e^-$, $B^+ \to K^0_S \pi^+ \mu^+ \mu^-$,
$B^0 \to K^+ \pi^- e^+ e^-$, and $B^0 \to K^+ \pi^- \mu^+ \mu^-$,
with $K^0_S \to \pi^+ \pi^-$.  Charge conjugation is implied here and throughout this note.
The submode $B^+ \to K^+ \pi^0 \mu^+ \mu^-$ is excluded from this analysis since Monte Carlo 
simulations indicated that it did not improve the results.  Events where the dilepton pair originated
from $J/\psi$ or $\psi(2S)$ decays are explicitly removed using selection criteria applied to $m_{\ell \ell}$.
The major background sources are semi-leptonic $B$ and $D$ decays, which are suppressed by applying optimized
cuts on bagged decision tree outputs.  The bagged decision trees utilize event shape variables, vertex information,
missing energy, and similar inputs to discriminate between signal and background events.  Training is carried out on Monte Carlo samples.  An additional background from $B^+ \to D \pi^+$, followed by $D \to K^* \pi$ along with 
$\pi \to \mu$ misidentification, is vetoed explicitly by rejecting $K^* \pi$ combinations near the $D$ mass.
After all selection criteria have been applied, the signal efficiency is typically about 15\%, although it varies by submode and $q^2$-bin.  

After event selection, a sequence of three maximum likelihood fits is performed to determine signal yields in each $q^2$-bin and to determine $F_L$ and $A_{FB}$.  The first fit determines the number of signal events in each $q^2$-bin.
It is performed in three dimensions: $m_{ES}$, $M(K \pi)$, and $\cal L$, where 
$m_{ES} = \sqrt{E^{*2}_{\rm beam} - p^{*2}_B}$,  $M(K \pi)$ is the mass of the 
$K^*$ candidate, and
$\cal L$ is a likelihood ratio formed from the output of the bagged decision trees used to separate signal events
from other $B$ meson decays.  $E^*_{\rm beam}$ and $p^*_B$ are the beam energy 
and momentum of the $B$ in the $\Upsilon(4S)$ frame (CM).

The subsequent fits 
are performed for the signal enriched region defined by $m_{ES} > 5.27 \, {\rm GeV}$.
In the second fit, a fourth dimension is added, $\cos \theta_K$, where $\theta_K$ is the angle between
the $K$ and the $B$ in the $K^*$ rest frame.  The yields determined from the first fit are fixed
in the second fit,
and the parameter $F_L$ is determined based on the equation:
\begin{equation}
{1 \over \Gamma} {d \Gamma \over d \cos \theta_K} = {3 \over 2} F_L \cos^2 \theta_K + 
{3 \over 4} (1 - F_L)(1 - \cos^2 \theta_K)
\end{equation}
In the final fit, $F_L$ is fixed at the result of the previous fit, and a fifth dimension is added, $\cos \theta_\ell$, where $\theta_\ell$ is the angle between the
$\ell^+$ ($\ell^-$) and the $B$ ($\Bbar$) in the dilepton rest frame.  The parameter $A_{FB}$ is determined from the fit based on the equation:
\begin{equation}
{1 \over \Gamma} {d \Gamma \over d \cos \theta_\ell} = {3 \over 4} F_L (1 - \cos^2 \theta_\ell) + 
{3 \over 8}(1 - F_L) (1 + \cos^2 \theta_\ell) + A_{FB} \cos \theta_\ell
\end{equation}
The fit methodology was validated using both Monte Carlo test experiments and by fits to 
$B \to K^* J/\psi$ and $B \to K^* \psi(2S)$ modes in data for which prior measurements exist.

\section{Results}

\begin{figure}[htb]
\begin{center}
\epsfig{file=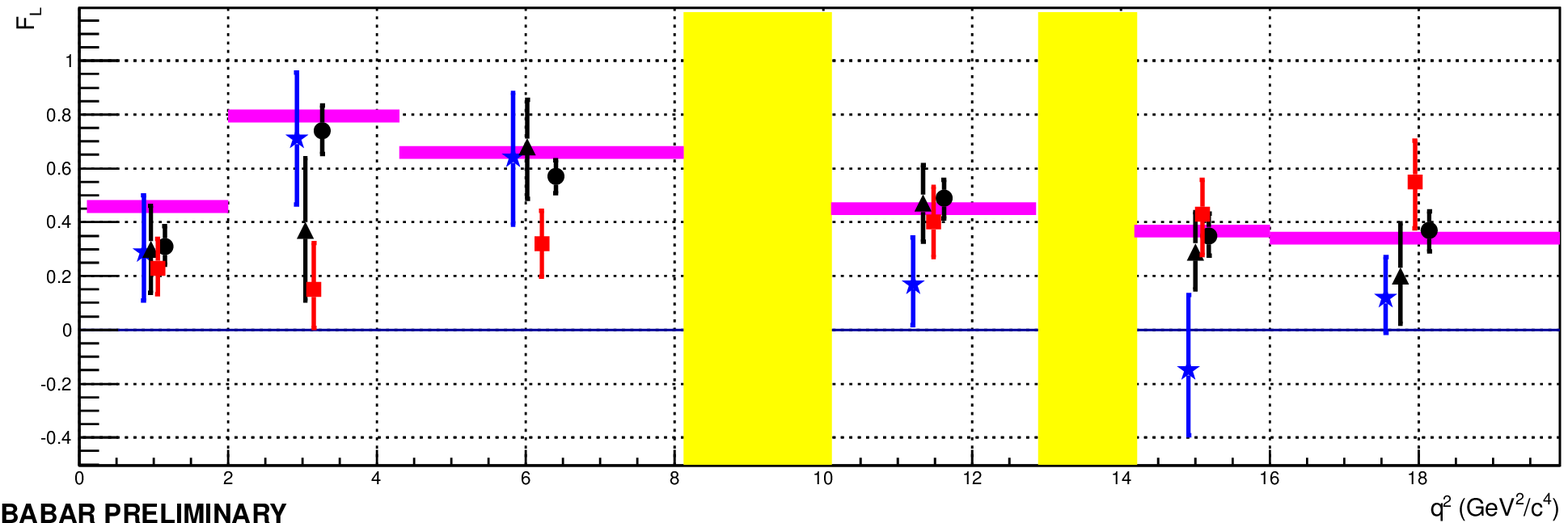,height=1.8in}
\epsfig{file=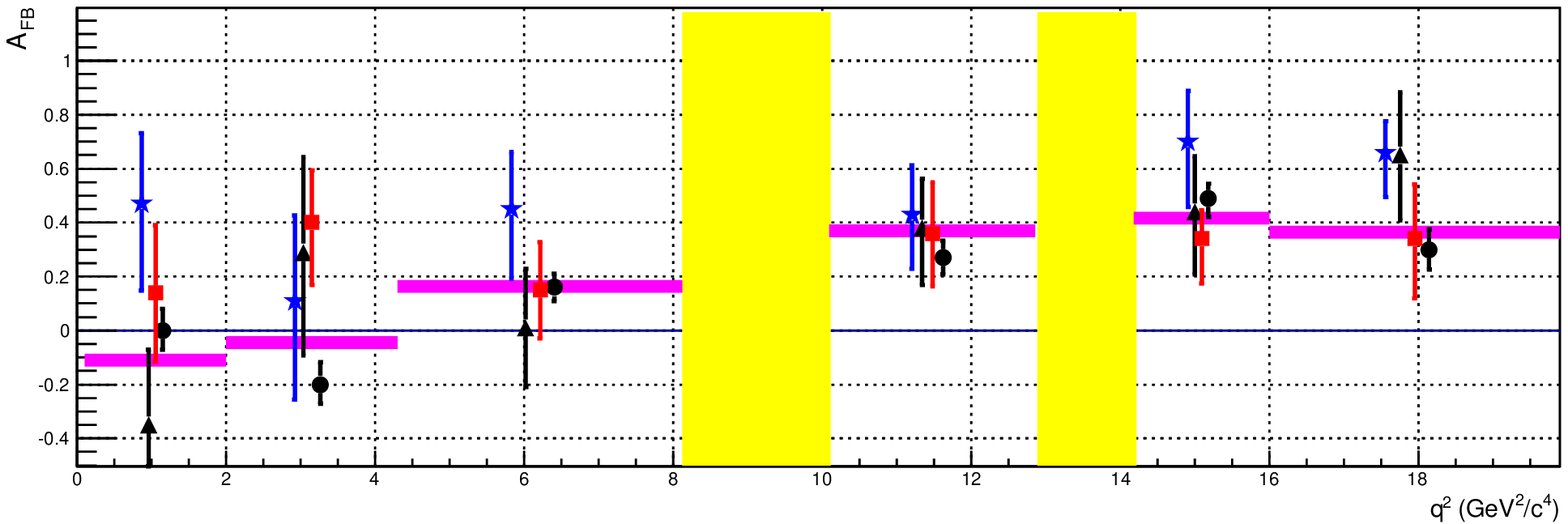,height=1.8in}
\caption{Results for $F_L$ (top) and $A_{FB}$ (bottom) versus $q^2$.  \babar\ preliminary results are
indicated by the solid red squares.  Results from other experiments are: Belle\cite{Belle} blue stars,
CDF\cite{CDFAFB} black triangles, and LHCb\cite{LHCb-Kstar} solid black dots.  Yellow shading indicates
the charmonium exclusion region.  The binned SM theory prediction \cite{Ali2} is shown in magenta.}
\label{fig:FLAFB}
\end{center}
\end{figure}

Preliminary \babar\ results for $F_L$ and $A_{FB}$ versus $q^2$ are given in Table~\ref{tab:FLAFB}.
While the dominant errors are statistical, systematic error estimates are included
in Table~\ref{tab:FLAFB}.  
Figure~\ref{fig:FLAFB} shows these results 
along with results reported thus far by other experiments.

%%%%%%%%%%%%%%%%%%%%%%%%%%%%%%%%%%%%%%%%%%%%%%%%%%%%%%%%%%%%%%%%%%%%%%%%%
%%
%%   use this format to include a LaTeX table  into your paper
%%
\begin{table}[htb]
\begin{center}
\begin{tabular}{l|cc}  
$q^2$ bin ($\rm GeV^2/c^4$) &  $F_L$  &  $A_{FB}$ \\ \hline
 0.1 - 2.0  &   $0.23^{+0.10}_{-0.09} \pm 0.04$   & $0.14^{+0.15}_{-0.16} \pm 0.20$       \\
 2.0 - 4.3 &  $0.15^{+0.17}_{-0.14} \pm 0.04$    & $0.40^{+0.18}_{-0.22} \pm 0.07$     \\ 
 4.3 - 8.1  &   $0.32 \pm 0.12 \pm 0.06$         & $0.15 \pm 0.16 \pm 0.08$      \\
 10.1 - 12.9 &  $0.40 \pm 0.12 \pm 0.06$         & $0.36^{+0.16}_{-0.17} \pm 0.10$     \\
14.2 - 16.00  &   $0.43^{+0.10}_{-0.13} \pm 0.09$ & $0.34^{+0.08}_{-0.15} \pm 0.07$  \\
 $>$ 16.00 &  $0.55^{+0.15}_{-0.17} \pm 0.03$      & $0.34^{+0.19}_{-0.21} \pm 0.07$       \\ \hline
1.00 - 6.00 & $0.25^{+0.09}_{-0.08} \pm 0.03$      & $0.17^{+0.12}_{-0.14} \pm 0.07$ \\ \hline
\end{tabular}
\caption{Preliminary \babar\ results for $F_L$ and $A_{FB}$ versus $q^2$.  In addition
$F_L$ and $A_{FB}$ measurements
are provided in a bin (1.00 - 6.00) that is sometimes used by theorists.
The first error is statistical and the second is systematic.}
\label{tab:FLAFB}
\end{center}
\end{table}
%%%%%%%%%%%%%%%%%%%%%%%%%%%%%%%%%%%%%%%%%%%%%%%%%%%%%%%%%%%%%%%%%%%%%%%%%%%

%%%%%%%%%%%%%%%%%%%%%%%%%%%%%%%%%%%%%%%%%%%%%%%%%%%%%%%%%%%%%%%%%%%%%%%%%
%%
%%   use this format to include an .eps figure into your paper
%%
%%%%%%%%%%%%%%%%%%%%%%%%%%%%%%%%%%%%%%%%%%%%%%%%%%%%%%%%%%%%%%%%%%%%%%%%%%%

\section{Conclusions}

Significant progress has been made in the study of $b \to s \ell^+ \ell^-$ related
$B$ decays.  Thus far no significant deviations from the SM have been observed.
Figure~\ref{fig:FLAFB} shows that the current results from all experiments are 
generally consistent with each other and with the SM expectations for $F_L$ and 
$A_{FB}$.  However, for meaningful comparisons of these angular measurements
with theory, much larger data sets are needed.
LHCb measurements will improve as additional data is collected.  Future
super-B factory results will also be important since a larger number
of final states will be accessible. 

This work was supported in part by Department of Energy contract DE-AC02-76SF00515.

%\bigskip
%I am grateful to Don Alfonso d'Alba for certain services essential to 
%this investigation.

%\def\Discussion{
%\setlength{\parskip}{0.3cm}\setlength{\parindent}{0.0cm}
%     \bigskip\bigskip      {\Large {\bf Discussion}} \bigskip}
%\def\speaker#1{{\bf #1:}\ }
%\def\endDiscussion{}

%\Discussion
%
%\speaker{D. Giovanni (University of Seville)}  My analysis indicates that the
%recovery of the two gentlemen is due simply to their embrace of the masculine
%principle and has nothing to do with magnetism at all.  Could you comment on 
%this?
%
%\speaker{Reggiano} Professor Giovanni has discussed this hypothesis in several
%forums, but, I do not believe there is anything in print.  I understand that
%he is spending his time in other pursuits.
%
%\speaker{D. Anna (University of Seville)}  In fact, my colleague Giovanni 
%has expressed opposite opinions on this question at various times, depending
%on the audience.  All of these testosterone-based theories are, of course,
%nonsense.
%
%\endDiscussion
 
\end{document}